# Total Ionizing Dose and Synergistic Effect of Magnetoresistive Random Access Memory

Xing-yao Zhang (张兴尧)[a), b) †], Qi Guo（郭旗）[a), b)], Yu-dong Li (李豫东)[a), b)],

Cheng-fa He (何承发)[a),b)], Lin Wen (文林)[a), b)],

a) Key Laboratory of Functional Materials and Devices for Special Environments, Xinjiang Technical Institute of Physics and Chemistry, Chinese Academy of Sciences, Urumqi 830011, China

b) Xinjiang Key Laboratory of Electronic Information Material and Device, Urumqi 830011, China

**Abstract:** Magetoresistive Random Access Memory (MRAM) was irradiated by $^{60}$Co γ-rays and electron beam. The test of synergistic effect was performed under additional magnetic field when irradiation. We analyzed Total Ionizing Dose (TID) and synergistic damage mechanism of MRAM. DC, AC and function parameters of the memory were tested in radiation and annealing by Very Large Scale Integrated circuit (VLSI) test system. The radiation sensitive parameters were obtained through analyzing the data. Because magnetic field imposed on MRAM when the test of synergistic effect, Shallow Trench Isolation (STI) leakage or Frenkel-Poole emission of synergistic effect was smaller than that of TID, and radiation damage of synergistic effect was lower than that of TID.

## 1. Introduction

MRAM is a new type nonvolatile memory, and it has many advantages, such as nonvolatility, unlimited endurance, fast write speed. Those make MRAM gotten much attention from commercial and military applications. MRAM uses Magnetic Tunnel Junction (MTJ) for data storage. Figure 1 shows the structure of an MRAM cell. Magnetic Tunnel Junction is formed from two ferromagnetic plates, which are Magnetic Free Layer and Magnetic Pinned Layer, each of them can hold a magnetic field. Tunnel barrier separated two ferromagnetic plates by a thin oxide layer. Bit line pass above the MTJ, and write line are below the MTJ. Magnetic Pinned Layer is a permanent magnet and set to a particular polarity. The magnetic field of Magnetic Free Layer is imposed by write line and bit line. The orientation of the magnetic field can be changed by the current direction of the bit line. If the two plates have the same polarity, the resistance of MTJ is low due to the magnetic tunnel effect, and this MTJ is considered to mean "1". While the two plates are of opposite polarity the resistance will be high and this means "0".

E-mail addresses: zxy@ms.xjb.ac.cn (Zhang Xing-yao)    Tel: 15909918289    Fax: 0991-3838957

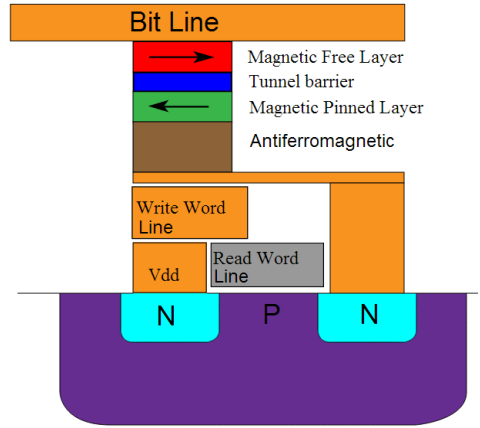

Figure 1 The structure of an MRAM cell

MTJ is written by magnetic fields, in other word, MRAM would be affected easily by strong magnetic field. There are strong magnetic fields from planet, Solar Flares and Spacecraft's magnetic torque of magnetorquer. Table 1 lists some values of magnetic field strengths found in the universe [1]. At the same time, Electrons and protons of Van Allen belts make total ionizing dose on MRAM in space. MRAM is imposed by synergistic effect, which is blend total ionizing dose with magnetic field in the real space environment.

Table 1 Values of magnetic field strengths

| Position | magnetic field /G |
|---|---|
| galaxy magnetic field | $10^{-5}$ |
| earth magnetic field | $\sim 5 \times 10^{-2}$ |
| solar magnetic field | $\sim 5 \times 10$ |
| Spacecraft's magnetic torque | $\sim 4 \times 10^2$ |
| Solar Flares | $\sim 5 \times 10^3$ |
| neutron star | $10^{11}$ |

The radiation damage of MRAM has been researched for many years since MRAM application appeared. MRAM has been analyzed radiation effect and mechanism for TID [2-6], SEL (Single Event Latchup) [2-4], SEU (Single Event Upsets) [3, 5, 7-9], SET (Single Event Transient) [7] and SEFI (Single Event Functional Interrupt) [7] test, but the test of synergistic effect has not been considered on MRAM. The purpose of this work is to analyze effect and mechanism for synergistic effect, which is radiation effect and magnetic field imposed on the MRAM at same time, will decrease or enhance the radiation damage of MRAM.

## 2. Test description and result

The MR25H10 (Everspin) memory is used. MRAM is organized as 131027×8 bits, with SPI interface up to 40 MHz clock rate.

Total ionizing dose test was done at the Xinjiang Technical Institute of Physics and Chemistry, Chinese Academy of Sciences, using $^{60}$Co γ source and electron beam from accelerator. The dose rate of different radiation source were very close, 0.78 Gy(Si)/s for $^{60}$Co γ source and 0.75Gy(Si)/s for electron beam, respectively. MRAM was in static test mode, which was put all Signal Pin in Vcc=3.3 V when TID test. Synergistic effect test and TID test were basically set to the same way, the only difference was additional magnetic field imposed the memory. The magnetic field was generated by a bulk of iron nickel permanent magnet, which was behind the MRAM. The magnetic intensity was tested by a Hand-held Gauss meter on the MRAM surface, and the magnetic intensity was about 250 G of horizontal orientation and 150 G of vertical orientation. It has ensured that all memories were functioning properly in that magnetic intensity. Electrical parameters were tested with the Verigy 93000. Data of the following parameters were recorded between radiation levels: standby current $I_{SB}$, write operation current $I_{DDW}$, read operation current $I_{DDR}$, Output high voltage $V_{OH}$, Output low voltage $V_{OL}$, Input leakage current $I_{LI}$, Output leakage current $I_{LO}$, SCK high time $t_{WH}$, the number of errors, and functional tests. Pattern of functional tests were read "55", write "00", read "00", write "FF", read "FF", write "55", read "55", sequentially.

Test of TID and Synergistic effect started reading errors at 400 Gy(Si) to different radiation source. But the number of errors when reading first "55" is different (see Table 2), the number of read errors of TID is more than that of synergistic effect. All memories restore functional within 24 hours at room temperature annealing.

Table 2 The number of read errors in test

| Numbers of errors | $^{60}$Co γ source | Electron beam |
|---|---|---|
| TID | 23840 | 101099 |
| Synergistic effect | 73 | 8864 |

The radiation sensitive parameters were standby current, write operation current and read operation current. Figure 2 and 3 shows the relationship among current, accumulated dose and annealing time in different radiation source. No matter what the radiation sources were, the current values increased with dose and reach maximal value until function failure, but the current values of total ionizing dose were bigger than those of synergistic effect at 400 Gy(Si).

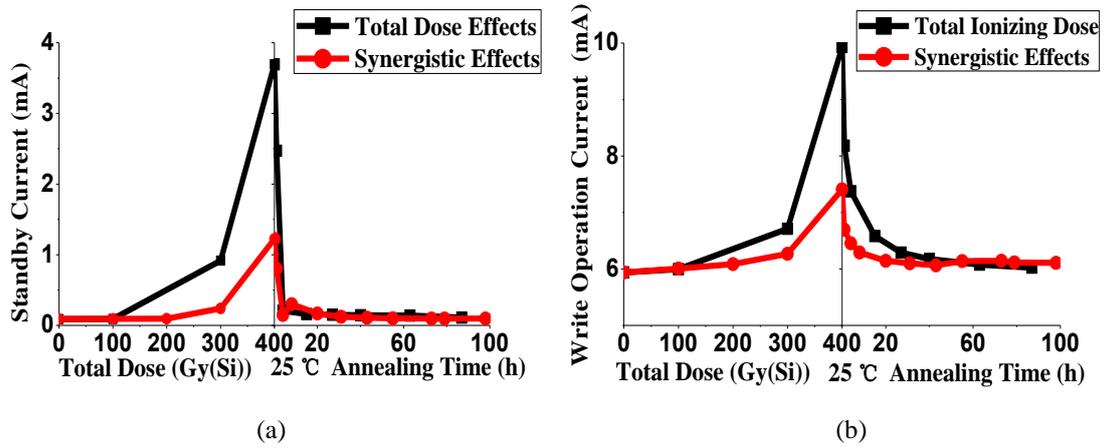

(a)　(b)

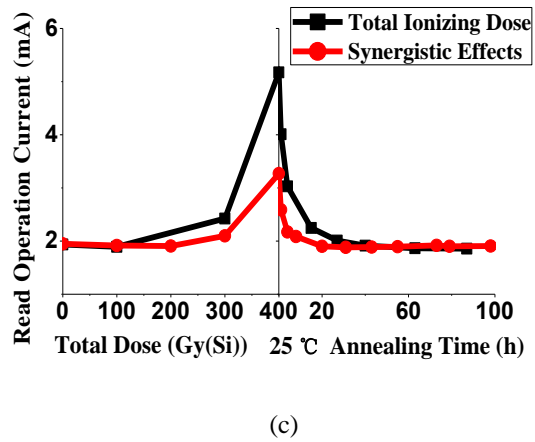

(c)

Figure 2 Currents change to $^{60}$Co γ source and annealing time, a for Standby current, b for Write operation current, c for Read operation current.

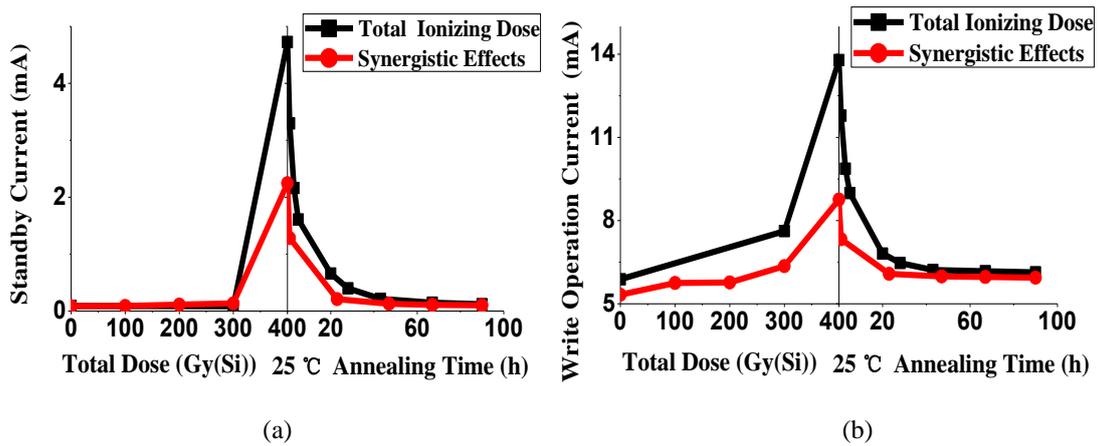

(a)　(b)

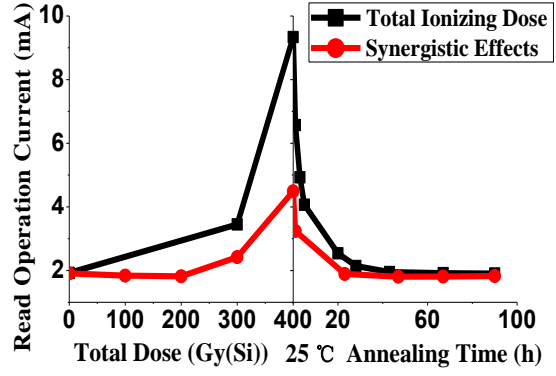

(c)

Figure 3 Currents change to electron beam and annealing time, a for Standby current, b for Write operation current, c for Read operation current.

## 3. Discussion

Because MTJ had good radiation toleration, it could work well under irradiation[10-12]. The performance of CMOS peripheral control circuit was deteriorated due to irradiation, and function of memory was fail immediately. Radiation produced a large number of oxide trapped charges, leading shallow trench isolation (STI) of MOS induced leakage[13,14], which was the major reason for deteriorating. Therefore the function failure of MRAM was resulted from CMOS peripheral control circuit rather than MTJ. MRAM began to restore function due to the disappeared of oxide trapped charges after annealing.

Figure 2 and 3 illustrated the currents change versus total dose at different radiation sources. The cause of currents raised was not only STI leakage but also MTJ leakage. Schottky emission was dominant part of MTJ leakage before radiation, the current density $J_{SE}$ was obtained in Schottky emission.

$$J_{SE} = A^* T^2 \exp[\frac{-q(\Phi_B - \Delta\Phi_{SE})}{kT}] \qquad (1)$$

$$\Delta\Phi_{SE} = \frac{1}{2}(\frac{q}{\pi\kappa_r\varepsilon_0})^{\frac{1}{2}} E^{\frac{1}{2}} \equiv \beta_S E^{\frac{1}{2}} \qquad (2)$$

Where $A^*$ is effective Richardson constant, $T$ is temperature, $q$ is elementary charge, $\Phi_B$ is the barrier height between the Fermi-level of the injecting metal, $\Delta\Phi_{SE}$ is due to image-force lowering[15,16], $\kappa_r$ is the dynamic dielectric constant, $\varepsilon_0$ is the permittivity in vacuum, $k$ is Boltzmann's constant, and

$E$ is electric field. Traps were engendered more and more in oxide layer of MTJ with the dose accumulating, with a positively applied bias to the electrode, electrons can tunnel first into the trap and then into the electrode, Frenkel-Poole emission became apparent[13,17]. The barrier height of Frenkel-Poole emission was the depth of the trap potential well, and the current density $J_{FPE}$ was obtained in Frenkel-Poole emission.

$$J_{FPE} = E\exp[\frac{-q(\Phi_T - \Delta\Phi_{FPE})}{kT}] \qquad (3)$$

$$\Delta\Phi_{FPE} = (\frac{q}{\pi\kappa_r\varepsilon_0})^{\frac{1}{2}}E^{\frac{1}{2}} \equiv \beta_{FPE}E^{\frac{1}{2}} = 2\beta_S E^{\frac{1}{2}} = 2\Delta\Phi_{SE} \qquad (4)$$

Denoting the barrier reduction by $\Delta\Phi_{FPE}$, the barrier reduction was larger than in the case of Schottky emission by a factor of 2, since the barrier lowering was twice as large due to the immobility of the positive charge [15]. Frenkel-Poole emission may dominate during exposure to ionizing irradiation, so MTJ leakage increased as Frenkel-Poole emission has increased.

The function of MRAM was failed at 400 Gy(Si) to TID and synergistic effect, but other parameters such as current and error numbers manifested the radiation damage of synergistic effect was less than that of TID. The experiment of vertical orientation magnetic field would explain that phenomenon. As TID, the electrons liberated by ionizing particle, and swept by electric field fast out of oxide layer. On the other hand, the holes were less mobile, and these become trapped, leading to the creation of oxide trapped charges finally. As synergistic effect, Magnetic field and electric field existed at the same time, part of electrons could not swept out of oxide immediately, but moved along a curved path because electric field was at right angle to the vertical orientation magnetic field at same time, as shown in figure 4. The net effect was that the part of electrons had an average "drift" in the direction of E×B[18], and figure 5 shows the drift motion in oxide. The electrons which drift in oxide recombined with holes, the number of holes transformed to oxide trapped charges of course is less than that of TID. As a result, STI leakage or Frenkel-Poole emission of synergistic effect was smaller than that of TID, so the parameters degeneration of MRAM showed the different appearance.

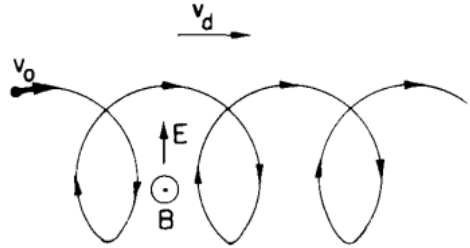

Figure 4 Path of an electron in crossed electric and magnetic field [18]

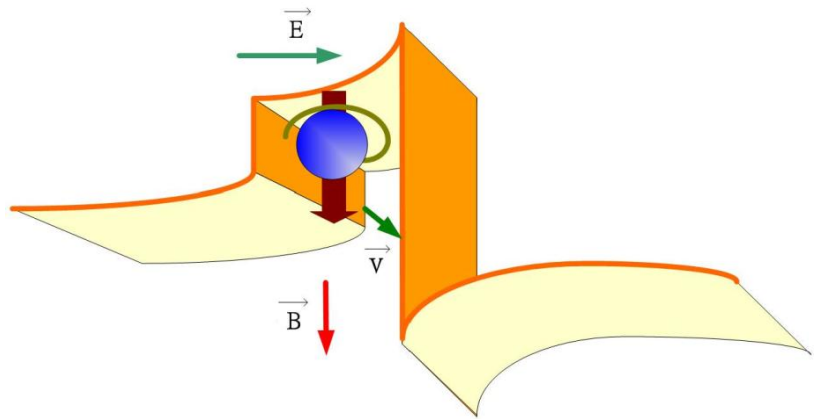

Figure 5 Electron drift in oxide

## 4. Conclusion

The function of MRAM would be failure at 400 Gy(Si) when TID and synergistic effect tests by different radiation sources. The reason of function failure was STI leakage lead to CMOS peripheral control circuit deteriorated. Standby current, read operation current and write operation current were changed in the test, so these parameters could be regarded as the radiation sensitive parameters. The radiation sensitive parameters manifested the radiation damage of synergistic effect is less than that of TID, because magnetic field make electrons drift in the oxide then recombine with holes, and lead to oxide trapped charges of synergistic effect less than those of TID. Hardening technique to tolerate TID is integrated onto SOI (silicon-on- insulator) CMOS supporting circuitry in the future.